\documentclass{article}

\PassOptionsToPackage{numbers,compress}{natbib}
\usepackage{arxiv}
\usepackage[utf8]{inputenc} % allow utf-8 input
\usepackage[T1]{fontenc}    % use 8-bit T1 fonts
\usepackage{hyperref}       % hyperlinks
\usepackage{url}            % simple URL typesetting
\usepackage{booktabs}       % professional-quality tables
\usepackage{amsfonts}       % blackboard math symbols
\usepackage{nicefrac}       % compact symbols for 1/2, etc.
\usepackage{microtype}      % microtypography
\usepackage{graphicx}
\usepackage{float}
\usepackage{placeins}
\usepackage{multirow}
\usepackage{caption}
\usepackage{listings}
\usepackage[table,xcdraw]{xcolor}
\usepackage{subfigure}
\usepackage{multirow}
\usepackage{amsmath} 
\usepackage{enumitem}
\usepackage{datetime2}
\usepackage{comment}
\usepackage{algorithm}
\usepackage{algpseudocode}
\usepackage{caption}

\title{How good are humans at detecting AI-generated images? Learnings from an experiment}
\date{} 					% Or removing it

\author{ {\hspace{1mm}Thomas Roca}\\
	Microsoft AI for Good Lab\\
	\And
	{\hspace{1mm}Anthony Cintron Roman}\\
	Microsoft AI for Good Lab\\
	\And
    {\hspace{1mm}Jehú Torres Vega}\\
	Microsoft AI for Good Lab\\
    \And
    {\hspace{1mm}Marcelo Duarte}\\
	Microsoft AI for Good Lab\\
    \And
    {\hspace{1mm} Pengce Wang}\\
	Microsoft AI for Good Lab\\
    \And
    {\hspace{1mm}Kevin White}\\
	Microsoft AI for Good Lab\\
     \And
    {\hspace{1mm}Amit Misra}\\
	Microsoft AI for Good Lab\\
    \And
    {\hspace{1mm} Juan Lavista Ferres}\\
	Microsoft AI for Good Lab\\
}

\renewcommand{\headeright}{May 2025}
\renewcommand{\undertitle}{ }

\hypersetup{
pdftitle={How good are humans at detecting AI-generated images},
pdfsubject={cs.CV},
pdfauthor={Thomas Roca, PhD},
pdfkeywords={Deepfake, Generative AI, experiment},
}

\begin{document}
\maketitle

\begin{abstract}
As AI-powered image generation improves, a key question is how well human beings can differentiate between "real" and AI-generated or modified images. Using data collected from the online game "Real or Not Quiz.", this study investigates how effectively people can distinguish AI-generated images from real ones. Participants viewed a randomized set of real and AI-generated images, aiming to identify their authenticity. Analysis of approximately 287,000 image evaluations by over 12,500 global participants revealed an overall success rate of only 62\%, indicating a modest ability, slightly above chance. Participants were most accurate with human portraits but struggled significantly with natural and urban landscapes. These results highlight the inherent challenge humans face in distinguishing AI-generated visual content, particularly images without obvious artifacts or stylistic cues. This study stresses the need for transparency tools, such as watermarks and robust AI detection tools to mitigate the risks of misinformation arising from AI-generated content

\begin{figure}[ht!]
    \centering
    \includegraphics[width=0.45\linewidth]{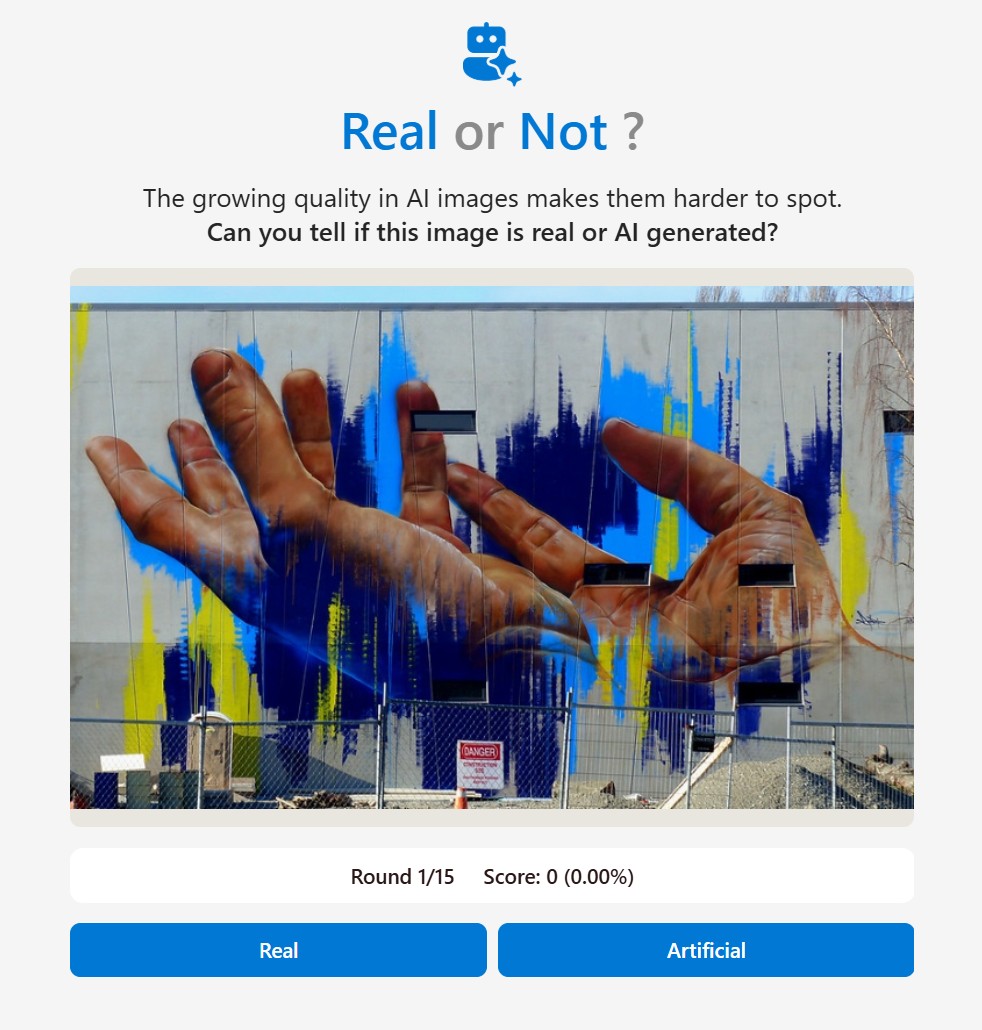}
    \caption{Example of image for user to evaluate - \url{https://www.realornotquiz.com/}}
    \label{fig:FIG1}
\end{figure}

\end{abstract}

\newpage
\section{Introduction}

In the past few years, generative AI made tremendous progress building on top of the diffusion architecture. Powered by cloud computing, the creation of realistic AI-generated audio, images and videos has been made accessible to everyone by simply using a text prompt, reference image or voice. Generative AI has been rapidly adopted by visual artists, communication agencies, influencers, etc., boosting their creativity and productivity. Unfortunately, it is also increasingly used by internet trolls, propaganda and disinformation campaigns and other malicious endeavors. Although image manipulation has a long history that predates even the digital era, the ease and the speed at which AI can produce content is unprecedented. 

Earlier this year, Microsoft launched a campaign called \textit{Combating the deceptive use of AI in elections}\footnote{https://news.microsoft.com/ai-deepfakes-elections/} to help with this issue. Training sessions were organized to raise awareness and prepare the public to tackle AI-generated misinformation during this year’s election cycles around the world. Our contribution to this effort was to show how hard it can be for people to spot AI-generated images of average quality. Previous research exploring this question either focused on Deepfake videos or when investigating AI-generated images have used generators now considered outdated. In both cases, they rely on a rather small number of human evaluations (100+).

To address these limitations, we created a game “the AI or not quiz”\footnote{https://aka.ms/aiornot} that shows a random sample of “real” or AI-generated images for users to guess. Our main objective is to understand how accurately people can tell AI-generated images apart from real ones, especially by examining the patterns or common mistakes people make\footnote{No PII data were collected}. 

\section{Methodology}
What sets our research apart is that it relies on a large sample of human evaluations through our online game. To date, the game has logged more than 82,500 distinct sessions. For this analysis we examine the cohort that played between 1 August and 8 August 2024: 12,500 participants around the world who completed the full game, which represent 17,340 unique games, producing roughly 287,000 image evaluations.

The image displayed in the game come from the collection of 350 copy-right free ‘real’ images and the generation of around 700 diffusion-based images using \textit{DallE-3}, \textit{Stable diffusion-3}, \textit{Stable diffusion XL}, \textit{Stable diffusion XL inpaintings}, \textit{Amazon Titan v1} and \textit{Midjourney v6}. We also added GAN\footnote{Generative Adversarial Network}-based fake faces.
Each user is shown one image at a time, for 15 rounds. The 15 images in a game are randomly selected from all images, both AI and real. The statistics we present in this paper are derived only from the results of completed games - in which all 15 images were evaluated by a user. 

The difficulty of such evaluation depends on how realistic AI-images appear and how ‘fake’, real images look. We decided that balancing the quality\footnote{That would be each image for a given generator would be representative of the average quality of this generator.} of AI-generated images across all generators would be difficult to achieve without preliminary assessment of a large pool of images. Instead, we tried for each generator to balance between easy (artifact clearly visible) and challenging images, but not necessarily in equal proportions per generator. For this reason, errors made by users are not indicative of the overall capability of a given generator to produce photorealistic images. Nevertheless, we hope to show some directions regarding the type of images that fool people more frequently. 

The game was designed to raise awareness about AI-generated images and how challenging it can be to spot them. We could have cherry-picked AI images that can fool most of the people, most of the time, but our goal was to provide a realistic panorama of the AI images people are likely to be exposed to, an “average” output of generative AI. However, our setting does not allow for a robust comparison between the generative AI models we use to generate our AI images as We do not use similar samples for each generator, nor balance photorealism across them.

\section{Results}
\subsection{Quantitative analysis}
\subsubsection{Can users spot AI-generated images? }
\begin{figure}[ht!]
\captionsetup{type=table}
    \centering
    \includegraphics[width=0.75\linewidth]{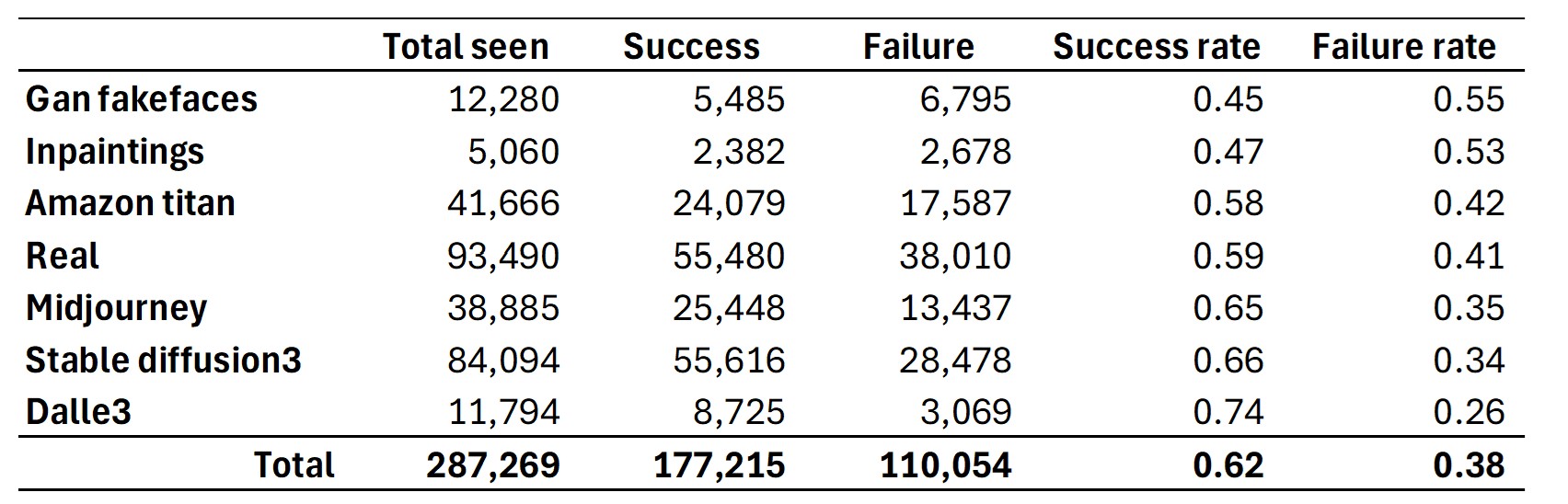}
    \caption{Success and failure rate}
    \label{fig:TABLE1}
\end{figure}

Table \ref{fig:TABLE1} shows the success and failure rate of our users. Out of the +287,000 images seen, around 110,000 were misidentified, corresponding to an overall success rate of 62\% (real and AI-generated). When looking specifically at AI-generated images, 193,779 were seen, of which 121,735 were accurately identified, resulting in a success rate of 63\%. These results suggest that, in this setting, our users’ odds at detecting AI images are only slightly higher than flipping a coin.

Figure \ref{fig:FIG2} below shows the density distribution of success scores for real images (blue) and for AI-generated ones (orange). Overall, it shows that there are only a few images that users consistently guess wrong. It also shows that among these more frequently misidentified images, a few real ones seem trickier, with a slightly flatter left tail.

\begin{figure}[ht!]
    \centering
    \includegraphics[width=0.5\linewidth]{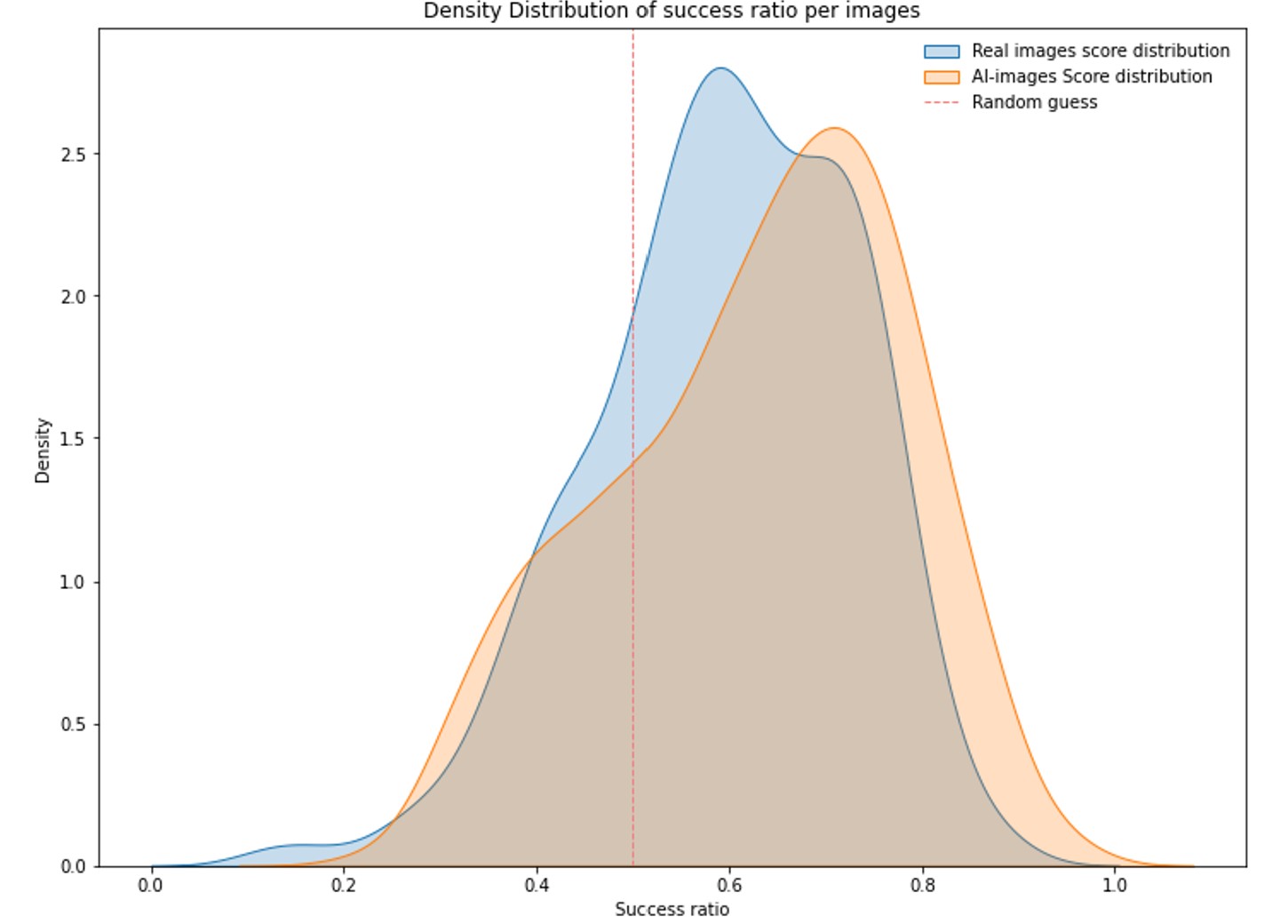}
    \caption{Density distribution of success ratio, real and AI-generated images}
    \label{fig:FIG2}
\end{figure}

\begin{figure}[ht!]
\captionsetup{type=table}
    \centering
    \includegraphics[width=0.65\linewidth]{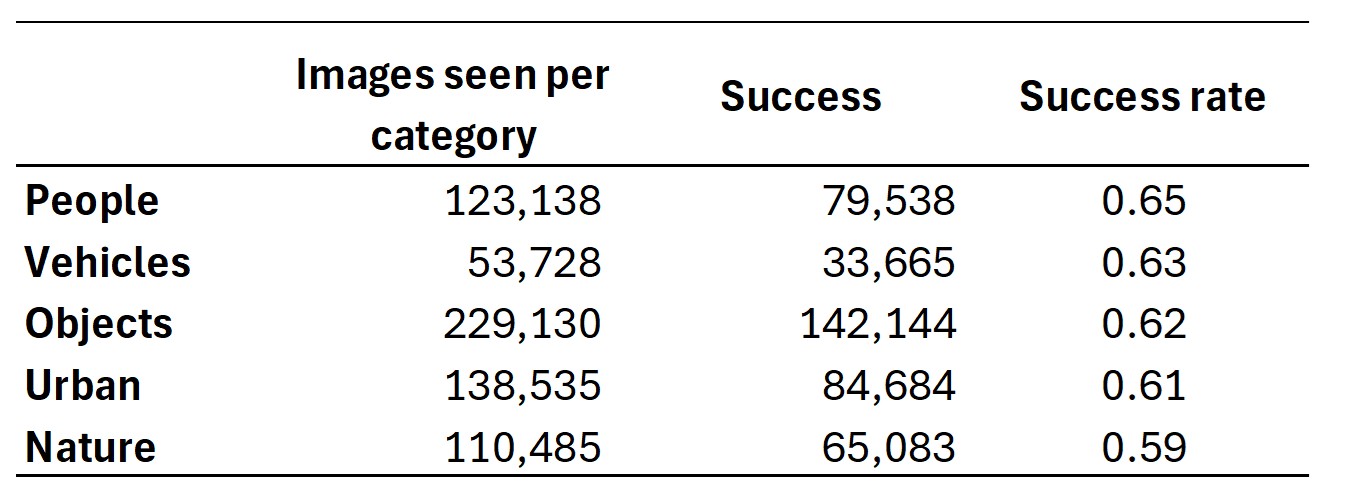}
    \caption{Success rate by image content}
    \label{fig:TABLE2}
\end{figure}

When looking at the success rate by image content - table \ref{fig:TABLE2} below -, we observe that users were the most successful when distinguishing between real people and AI generated ones. The categories they failed the most are natural and urban landscapes.

We suggest these results could derive from humans’ high ability to identify faces - see Oruc, I. et al. (2019)\cite{Oruc2019} and Wilmer, J.B et al. (2010)\cite{Wilmer2010} – and that abnormalities are thus easier to detect compared to landscapes. 
Overall, these results are in line with the literature. Zeyu Lu et. al. (2023)\cite{Lu2023} shows a similar failure rate and describes how performance varies depending on the image (higher accuracy observed on images depicting people and lower accuracy for objects and landscapes, see figure \ref{fig:FIG3} below).
Evaluating people’s ability to detect deepfake videos, Somoray, K (2023)\cite{Somoray2023} and Nils C. Köbis et. al. (2021)\cite{Kobis2021} also observe a success rate close to 60\%. 

\begin{figure}[ht!]
    \centering
    \includegraphics[width=0.7\linewidth]{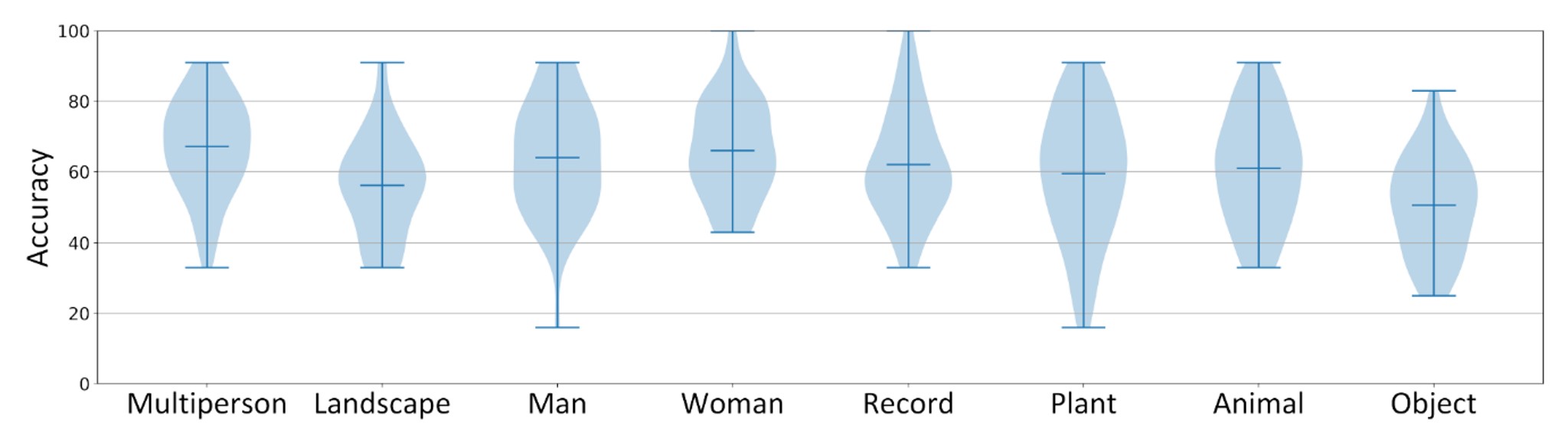}
    \caption{Accuracy by image category, Zeyu Lu and al. (2023)}
    \label{fig:FIG3}
\end{figure}

These numbers are averages and are highly dependent on the visual quality of the AI generated image. As previously mentioned, it would be easy to cherry pick AI-images capable of fooling people. The most recent models like \textit{Flux} -  from \textit{Black Forest lab} -  can generate images that look like amateur photography. In that case, the generator would not optimize for aesthetics, but the specific prompt would guide generation towards ‘amateur’ images. It is common practice for diffusion models to be evaluated and fine-tuned to generate images based on human feedback for quality and aesthetics (see Yang, K, 2023)\cite{Yang2023}. Most diffusion-based models optimize for aesthetic/quality images, rather than amateur /non-professional looking ones. We assume that people have started to perceive or associate AI-generated images with a certain aesthetic and style, often over-refined and mimicking studio-quality images – not that they cannot produce other styles of images.

\begin{figure}[ht!]
    \centering
    \includegraphics[width=0.75\linewidth]{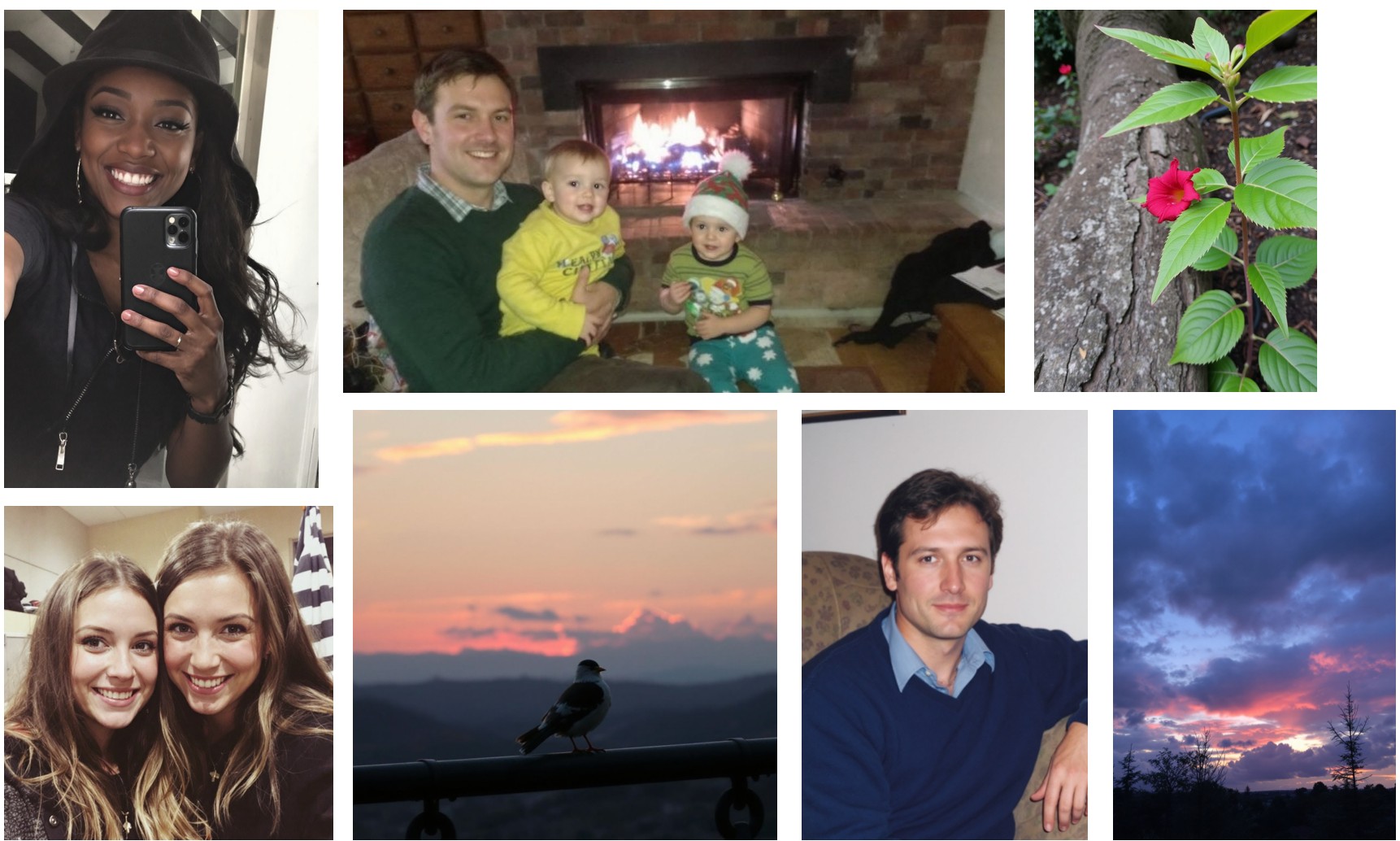}
    \caption{AI images (Flux pro) generated using prompts seeking amateur style. These images are not used in our game.}
    \label{fig:FIG4}
\end{figure}

\subsubsection{Can “image quality” assessment inform users’ success?}
\begin{figure}[ht!]
\captionsetup{type=table}
    \centering
    \includegraphics[width=0.75\linewidth]{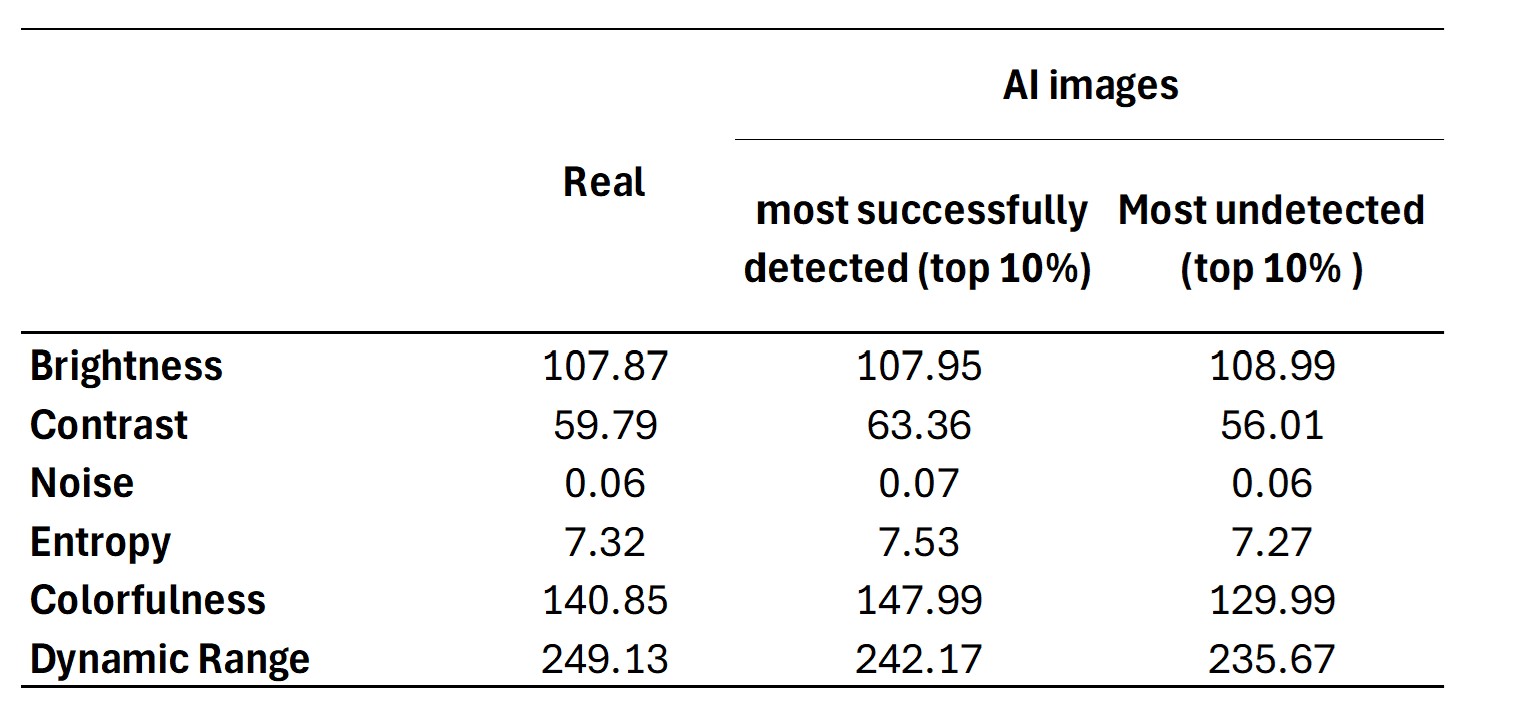}
    \caption{Image Quality Assessment}
    \label{fig:TABLE3}
\end{figure}

To better understand whether image properties influence users’ success, we conducted a brief Image Quality Assessment (IQA). Table \ref{fig:TABLE3} shows averages for a set of metrics for real images, and AI images. For the AI images, we selected the top 10\% of images that users correctly identified most often and the 10\% that users misidentified the most. For all the metrics, dynamic range excepted, a higher value appears to help users distinguish AI images from real ones. Although consistent across these metrics, it is hard to evaluate if the differences in these values are noticeable for the human eye. To understand success and failure better, we conduct in section 3.2, a qualitative assessment of the images users fail and succeed to identify the most.

\begin{figure}[ht!]
    \centering
    \includegraphics[width=0.65\linewidth]{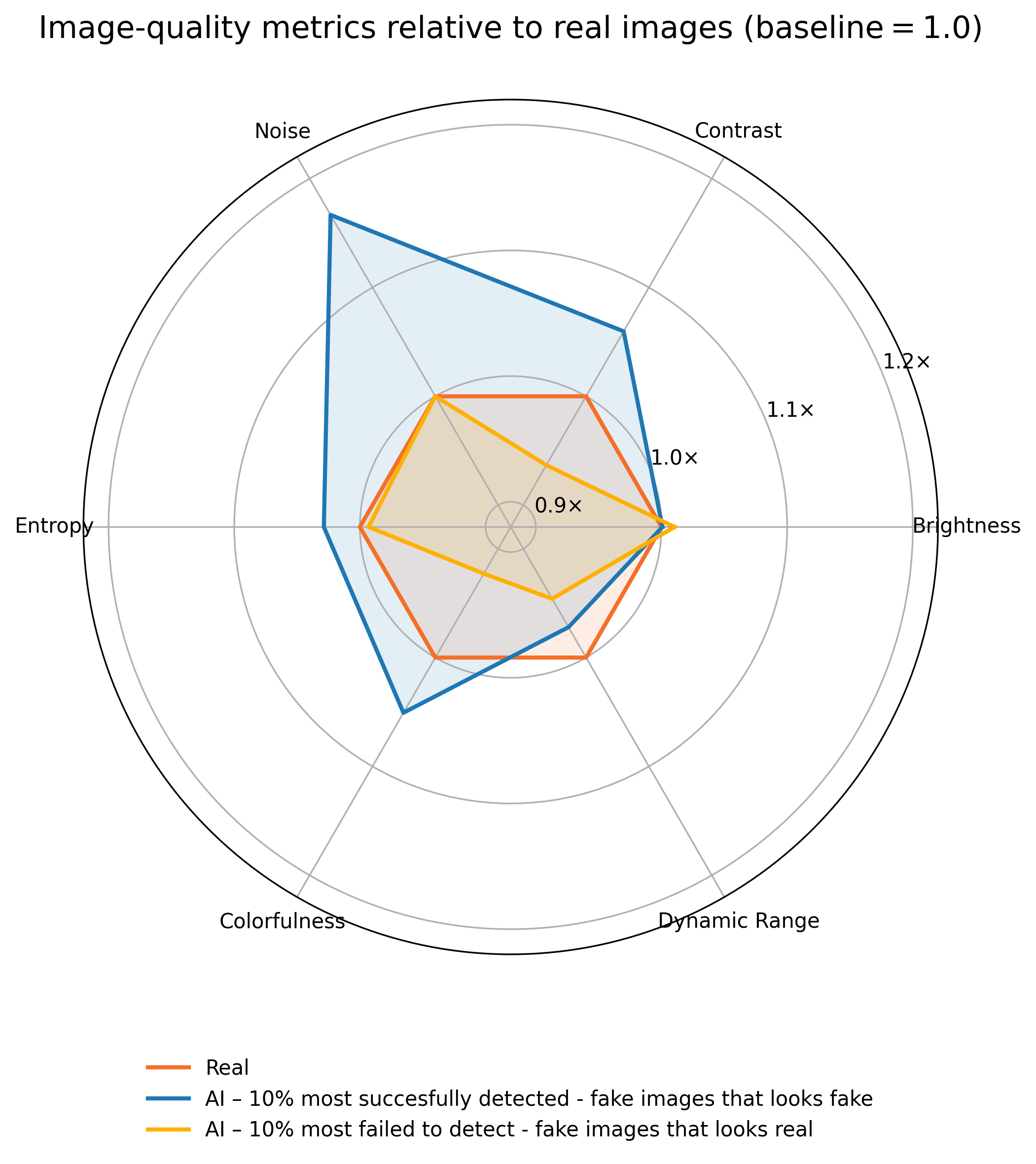}
    \caption{Image Quality Assessment}
    \label{fig:FIG5}
\end{figure}

\subsubsection{Which generator fooled people the most?}
As previously mentioned, this game was not designed to scientifically evaluate photo realism across several generative AI models. Our selection of images does not reflect the quality of the underlying models, and not all models are equally represented in our AI-generated image sample. Nevertheless, we can observe that two generation systems have a success rate below 50\%: Generative Adversarial Network (GAN) depicting human faces and Inpaintings. 

\begin{figure}[ht!]
    \centering
    \includegraphics[width=0.75\linewidth]{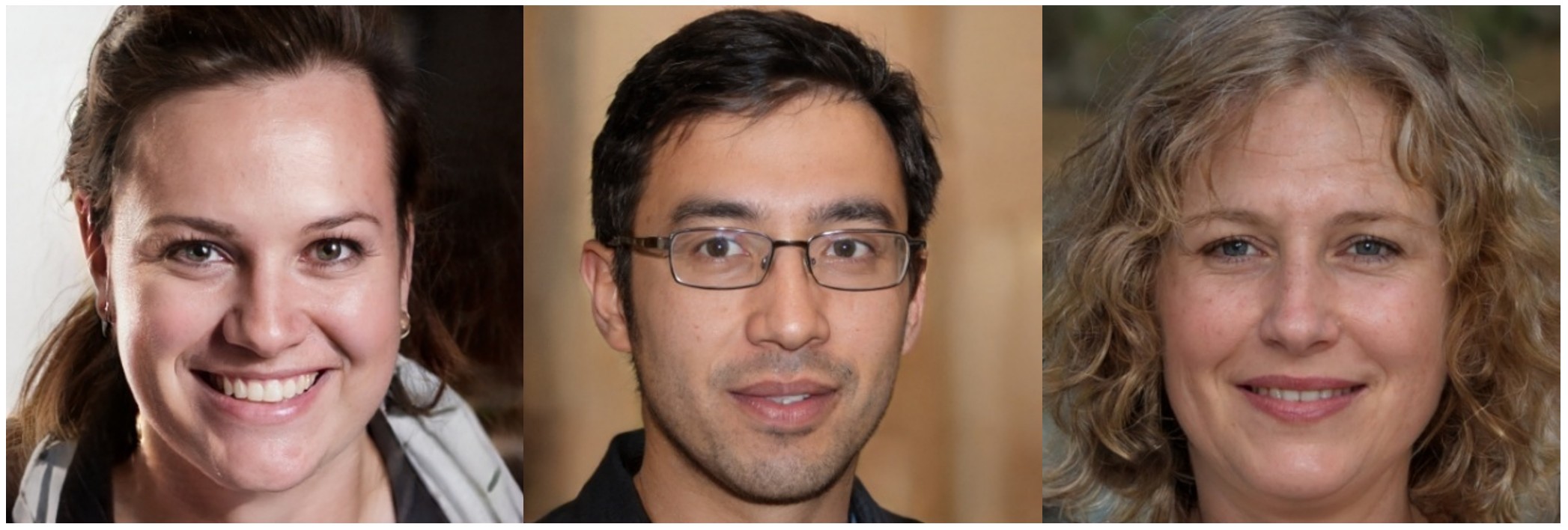}
    \caption{GANs fake face examples}
    \label{fig:IMG3}
\end{figure}

Although GANs are now 10 years old - see Goodfellow and al 2014 \cite{goodfellow2014generativeadversarialnetworks} - and represent the previous wave of generative AI models, they are known for the quality of the images they produce - see Figure \ref{fig:IMG3}. However this quality comes with a tradeoff, they are hard to train and do not generalize beyond the examples they have seen during training which greatly limits their usability. When it comes to face depiction though, these models are still very commonly used in video deepfake generation systems. The images they produce look very similar to “amateur” images – they were trained on similar images – yet most of these models were not trained to reproduce the studio-like aesthetic usually produced by models like \textit{Midjourney}. Again, we should not assume that a model architecture is responsible for the aesthetic of its output, the training data is. The model architecture only determines how successful a model is at mimicking a training set.

Inpainting is a distinct category, it is a technique rather than a model. Inpainting consists of replacing a given element in a picture with another AI-generated one. In our case, inpaintings were generated using \textit{Stable Diffusion XL}, but we could have used other diffusion-based models.

\begin{figure}[ht!]
    \centering
    \includegraphics[width=0.75\linewidth]{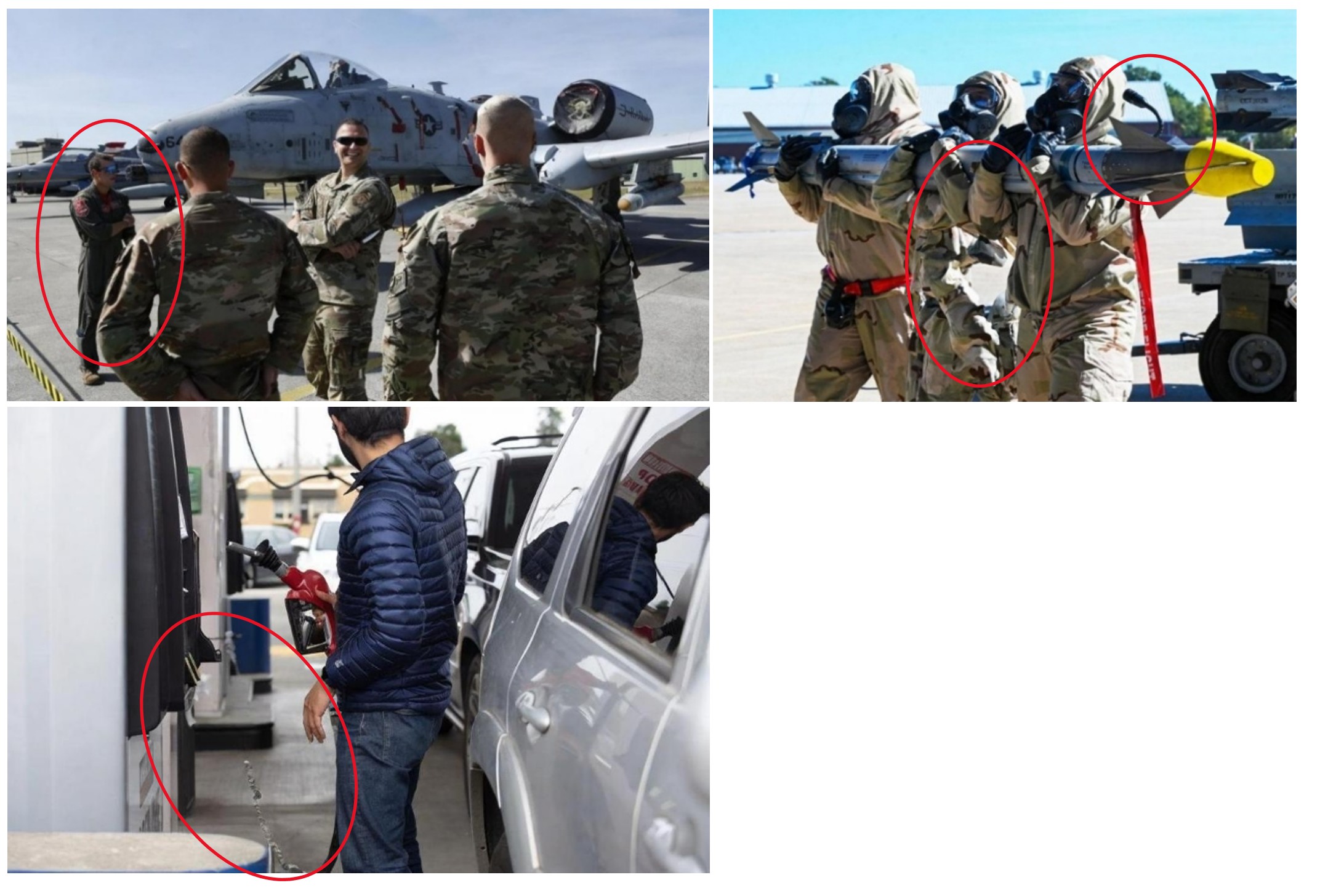}
    \caption{Examples of inpaintings gone wrong, the red circles indicate regions that are AI generated }
    \label{fig:IMG2}
\end{figure}

Our dataset of inpaintings were generated by randomly selecting an area/object in a picture using image segmentation. A multi-modal model was used to describe the object extracted, and then the description was used in a prompt to replicate and replace it using generative AI. The outcome is an image that is mostly ‘real’ pixels but contains an AI generated element. This technique makes it hard to identify forgery and can be used to generate fake images in disinformation campaigns, for example replacing people, adding objects, etc. We suggest it is important that the public is aware that this technique exists.
When it comes to other generative AI models, \textit{DallE-3}, \textit{Midjourney} and \textit{Stable Diffusion }AI-images seem easier to correctly identify. These 3 models are the most notorious ones and images produced by them have been widely published. We assume these images have a characteristic style and that the public is now used to seeing and recognizing.

\subsection{Qualitative analysis}
\begin{figure}[ht!]
    \centering
    \includegraphics[width=0.95\linewidth]{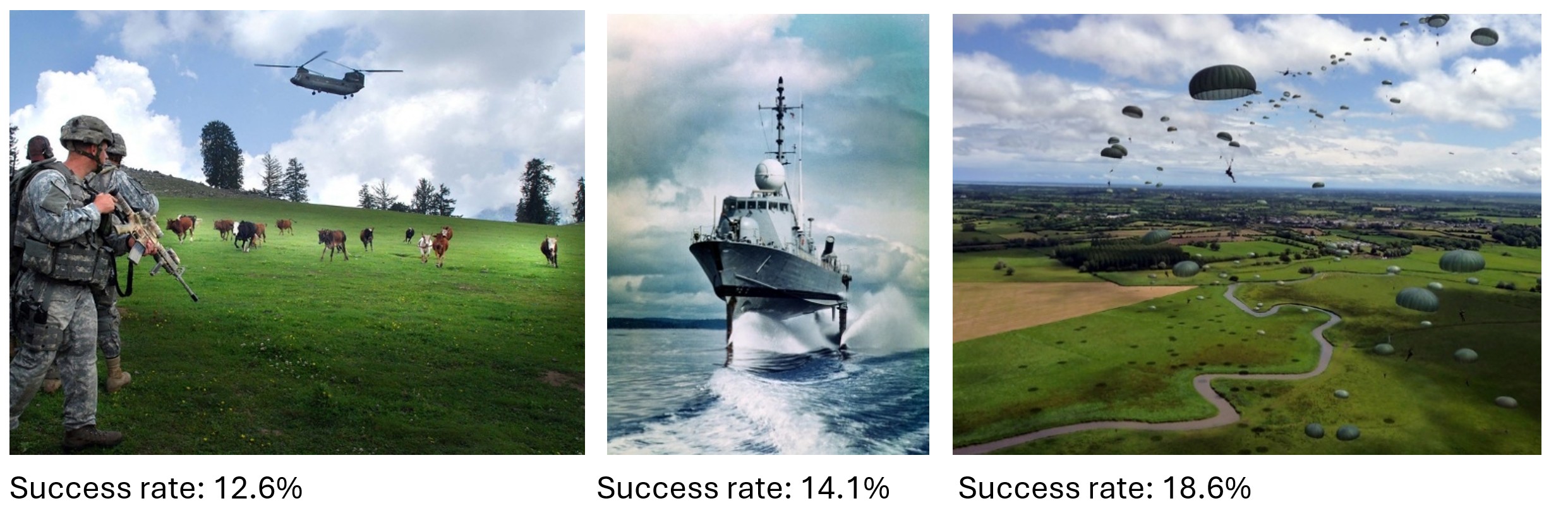}
    \caption{Real images with low success rate}
    \label{fig:IMG4}
\end{figure}

Out of the 1,000+ images that can be displayed in our game, only 3 images have a success rate lower than 20\%. The 3 of them are real images - figure \ref{fig:IMG4}. Among all images with a success rate under 25\%, only 2 are diffusion-based.

The real images shown in figure \ref{fig:IMG4} share similar aesthetics with the images produced by generative AI models. The light and color can indeed seem unnatural or uncommon, as can the scenes depicted. These 3 images are from the United States national archives. The one on the left has the lowest success rate (12.6\%). It shows U.S. troops in Afghanistan, the lights, the colors, the rotor blades of the chopper that seem frozen due to the camera’s fast shutter’s speed, etc., make this image particularly tricky to evaluate for people.  

\begin{figure}[ht!]
    \centering
    \includegraphics[width=0.95\linewidth]{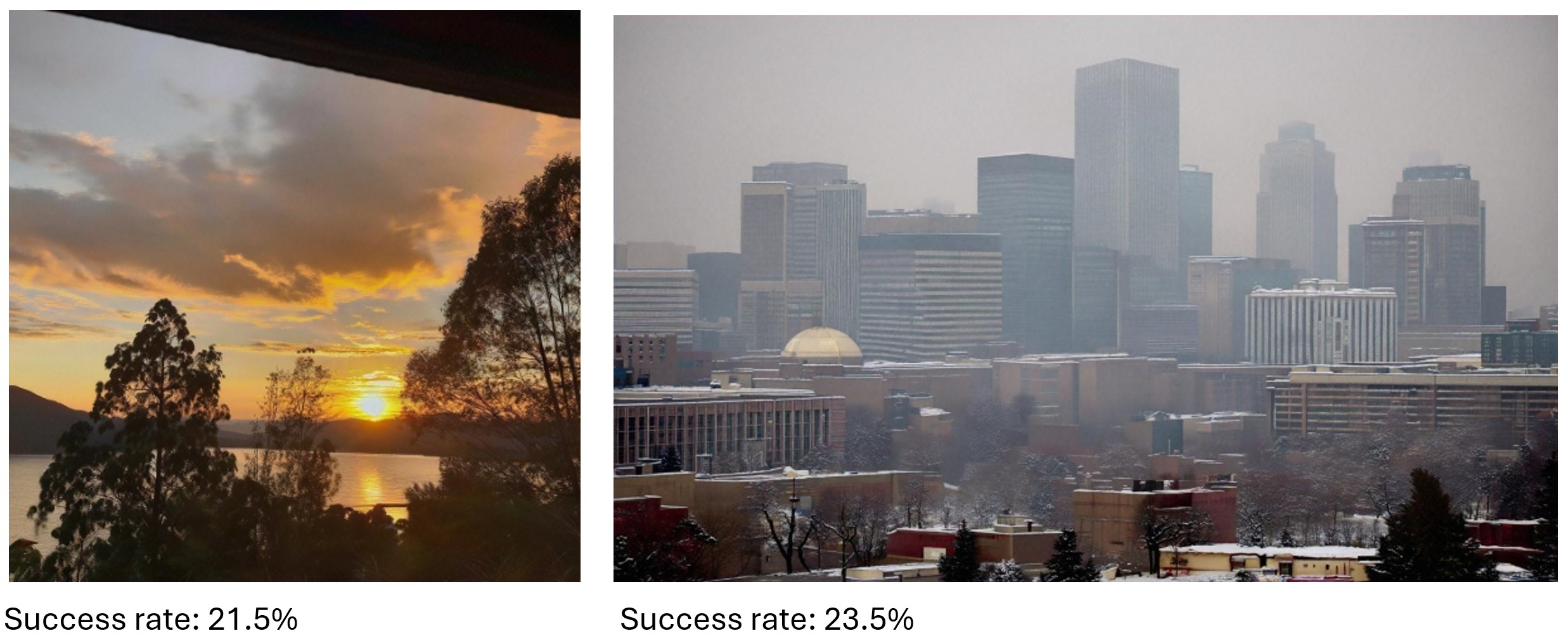}
    \caption{AI-generated images with low success rate}
    \label{fig:IMG5}
\end{figure}

The two AI-generated images with a success rate under 25\% - Figure \ref{fig:IMG5}-  were generated by Amazon's \textit{Titan v1} model, using guided generation – an image was used together with a prompt to generate the image. This technique allows a model to take image composition and style into account, producing a more photorealistic outcome.

\section{How do users compare to AI detection tools?}
Not all AI-generated images seen by the public will display a watermark or contain a digital signature that would allow viewers to rapidly identify them as AI-generated. AI detection tools can be helpful in distinguishing a “real” image from an AI-generated one. The success rate of such detectors can vary and the features that help a machine learning model to spot AI images are not totally understood. But most AI detectors have a success rate significantly higher than people. The AI detector our team is developing has a success rate superior to 95\% on both real and AI generated images - see table \ref{fig:TABLE4} below. AI detectors’ success rate is also consistent across image categories. 

\begin{figure}[ht!]
\captionsetup{type=table}
    \centering
    \includegraphics[width=0.6\linewidth]{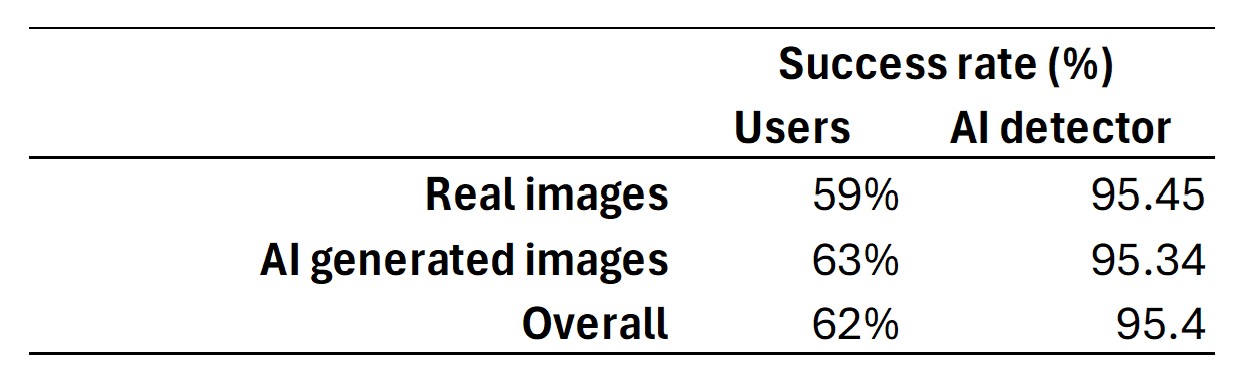}
    \caption{Our AI detection tool success rate}
    \label{fig:TABLE4}
\end{figure}

\section{Conclusion}
Although we did not cherry-pick images to fool people, our users’ ability to distinguish AI-images from real ones is just slightly better than flipping a coin. Only a handful of images have a significant failure rate, and most of the 1,000 pictures we selected appear to be equally challenging for our users. We suggest that most generative AI models nowadays can produce photorealistic images without visible defects. We also suggest that people are now used to seeing the output of generative AI models such as \textit{Dall-E}, \textit{Midjourney} or \textit{Stable Diffusion} and that they can recognize the ‘average’ aesthetic of them, aesthetic driven by the model builders and training dataset rather than capability of the model to produce images that could deceive viewers. AI images that fool users the most are images that look realistic but not professional. AI images depicting landscapes and objects seem harder to detect by users, likely due to the inherent ability of people to identify human faces.

Generative AI is evolving fast and new or updated generators are unveiled frequently, showing even more realistic output. It is fair to assume that our results likely overestimate nowadays people's ability to distinguish AI-generated images from real ones.
Finally, this experiment makes a strong case for more transparency when it comes to generative AI. Content credentials and watermarks are necessary to inform the public about the nature of the media they consume. In instances where digital signatures are not available, AI detection tools can be used, as they are significantly more reliable than humans at identifying AI-images - keeping in mind that them too, will make mistakes.

\newpage
\bibliographystyle{plain}
\bibliography{citations}
\appendix

\newpage

\end{document}